\documentclass[sigconf,nonacm]{acmart}

\setcopyright{none}
\acmConference[KDD Cup 2026 UniRec Workshop]
{KDD Cup 2026 Tencent UniRec Challenge Workshop}
{August 12, 2026}
{Jeju, Korea}

\acmYear{2026}
\acmDOI{}
\acmISBN{}

\AtBeginDocument{%
  }

\newcommand\blfootnote[1]{%
  \begingroup
  \renewcommand\thefootnote{}%
  \footnotetext{#1}%
  \addtocounter{footnote}{-1}%
  \endgroup
}

\begin{document}

\title[Dense Feature Representation over Sequence Modeling]{Dense Feature Representation over Sequence Modeling:\\
A Solution to the KDD Cup 2026 UniRec Challenge}

\author{Yi Zhang}
\affiliation{%
  \institution{Z Lab}
  \city{Chengdu}
  \state{Sichuan}
  \country{China}}
\email{ian.zhang.cn@protonmail.com}

\author{Weiliang Ji}
\affiliation{%
  \institution{Z Lab}
  \city{Chengdu}
  \state{Sichuan}
  \country{China}}
\email{jwl0711@mail.ustc.edu.cn}

\renewcommand{\shortauthors}{Zhang and Ji}

\begin{abstract}
We describe our 10th-place solution to the KDD Cup 2026 Tencent UniRec Challenge, industrial
click-to-conversion (CVR) prediction over 34.82M records, and we ask which mechanisms actually move held-out
AUC. Starting from the official PCVRHyFormer baseline, a 15-step single-variable chain raises test AUC from
0.813237 to 0.827816, and our final submission reaches 0.828535. A leave-one-out ablation from the full model
attributes the gain: removing the dense-feature representation stack costs 0.0095 AUC and removing the
orthogonalized optimizer costs 0.0028, while no sequence-modeling component (merged single-stream
backbone, polarity channel, auxiliary head, per-token FFN) costs more than 0.0005, within or adjacent to a
$\pm$0.0004 seed band. We also
report a generalization hazard: the row-group train/validation split shares one time window, so validation
AUC overstates the leaderboard by about 0.014; anti-memorization and high-cardinality-ID changes even invert
sign against it, a divergence that traces to dump-to-dump distribution shift and survives a time-ordered
re-split. Dense representation and optimization, not finer sequence modeling, drive CVR AUC
at this scale, and verdicts must come from the held-out leaderboard.
\end{abstract}


\maketitle
\blfootnote{\scriptsize
KDD Cup 2026 Tencent UniRec Challenge Workshop, August 12, 2026, Jeju, Korea.\\
Competition website: \url{https://algo.qq.com/}.
}

\section{Introduction}
\label{sec:intro}

Conversion-rate (CVR) prediction estimates whether a click converts into a purchase, and it governs
revenue allocation in large-scale recommendation and advertising systems. The KDD Cup 2026 Tencent
UniRec Challenge~\cite{unirecchallenge2026} poses this task at industrial scale: 34.82M click records,
a single model that jointly encodes sequential user behavior and multi-field non-sequential features,
and a held-out leaderboard scored by test AUC.

Recent unified backbones leave two structural questions open. First, HyFormer~\cite{hyformer},
OneTrans~\cite{onetrans}, and MixFormer~\cite{mixformer} combine sequence modeling, feature interaction,
and training tricks in one architecture, so which mechanism actually moves
held-out AUC at this scale stays unmeasured. Second, the standard train/validation split orders rows by
row group rather than by time, so validation AUC and the leaderboard measure different distributions, a
gap wide enough to invert the sign of some changes.

We answer both questions with an ablation study against the leaderboard. Starting from our strongest model,
each variant removes exactly one component (a leave-one-out ablation), and we measure the resulting drop
in leaderboard AUC, which isolates each mechanism's contribution to generalization rather than to
in-distribution validation.

Our solution extends the official PCVRHyFormer baseline through a 15-step single-variable chain. Each step
changes exactly one mechanism. The chain adds a dense-feature representation stack (field splitting, log1p,
normalization), a cross-domain merged single-stream sequence backbone, a target-conditioned polarity channel
that separates positive from negative behavior~\cite{tapf}, a sparse anti-memorization regularizer, and an
orthogonalized optimizer~\cite{amuse,muon}.

Our contributions are empirical:
\begin{itemize}
  \item A single-variable upgrade chain that raises test AUC from the official baseline 0.813237 to
  0.827816 ($+$0.014579); our final submission reaches 0.828535 and places 10th (\S\ref{sec:exp}).
  \item A leave-one-out attribution: dense representation (0.0095 AUC when removed) and optimizer choice
  (0.0028) account for nearly all of the gain, while every sequence-modeling component stays within or
  adjacent to a $\pm$0.0004 seed band (\S\ref{sec:exp}).
  \item Evidence that the shared-time-window split inflates validation AUC by about 0.014, and that
  anti-memorization and high-cardinality-ID changes invert sign against the leaderboard through dump-to-dump
  distribution shift that a time-ordered re-split does not fix (\S\ref{sec:exp}).
  \item Negative results reported as first-class findings: feature-crossing and category target-attention
  mechanisms screened out on validation, and hand-built user-behavior features with no measured test gain
  (\S\ref{sec:discussion}).
\end{itemize}

\section{Related Work}
\label{sec:related}
\textbf{Unified backbones.} One line fuses feature interaction and sequence modeling in a single
architecture: HyFormer~\cite{hyformer}, OneTrans~\cite{onetrans}, MixFormer~\cite{mixformer}, and
InterFormer~\cite{interformer}; several of them benchmark against the token-mixing feature-interaction
network RankMixer~\cite{rankmixer}. These architectures ship many mechanisms at once, so each mechanism's
marginal effect on a time-shifted held-out set stays open.

\textbf{Scaling and long-sequence modeling.} A second line scales recommendation models along parameters and
history length: Wukong~\cite{wukong}, HSTU~\cite{hstu}, LONGER~\cite{longer}, and unified scaling-law
analyses~\cite{kunlun}. These results measure gains on in-distribution metrics; whether width or
sequence-length scaling transfers to a held-out dump under a non-temporal split stays unexamined.

\textbf{Sequence content and optimization.} A third line changes what the model consumes or how it trains
rather than the backbone: TAPF~\cite{tapf} interleaves negative behavior into the sequence, and orthogonalized
optimizers such as Muon~\cite{muon} and AMUSE~\cite{amuse} reshape the update. Each is validated in isolation
or on other tasks, so its marginal contribution atop a strong dense-and-optimizer stack, on industrial CVR,
is unknown.

\textbf{Offline evaluation under temporal shift.} Conversion labels arrive with delay, which censors recent
examples~\cite{chapelle2014delayed}; the split strategy changes which models win
offline~\cite{meng2020splitting}; and splits that ignore the timeline leak future information into
evaluation~\cite{ji2023leakage}. We add a quantified case from a live industrial leaderboard: an iid
row-group split inflates validation AUC by 0.014 and inverts the sign of regularization and capacity changes
(\S\ref{sec:exp}).

\textbf{Positioning.} We do not propose a new backbone. We take the official PCVRHyFormer baseline, add
mechanisms from these three lines one at a time and, through leave-one-out ablation against a
held-out leaderboard, report which move test AUC and which do not, and how a non-temporal split distorts that
measurement.

\section{Method}
\label{sec:method}

\subsection{Architecture overview}
\begin{figure*}[t]
  \centering
  \includegraphics[width=\textwidth]{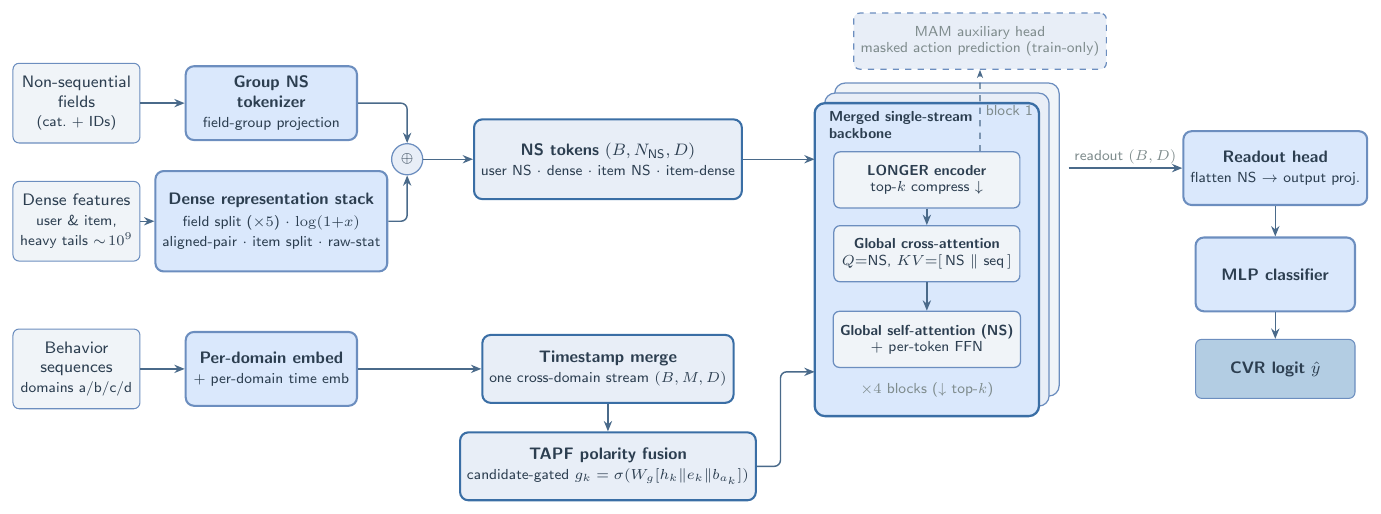}
  \caption{The solution pipeline. Three input modalities (non-sequential fields, dense user and item
  features, per-domain behavior sequences) tokenize into a shared space. Non-sequential and dense tokens
  form the NS token set; the behavior domains merge into one timestamp-ordered stream that a
  target-conditioned TAPF polarity gate fuses with the candidate item. The merged single-stream backbone
  stacks four blocks, each compressing the sequence through decreasing top-$k$ retention (LONGER encoder)
  and mixing it with the NS tokens through global cross- and self-attention with a per-token FFN; the
  readout flattens the NS tokens into a CVR logit. The MAM auxiliary head (dashed, training only) predicts
  masked actions from the first block. The dense-representation stack and the optimizer, not the sequence
  backbone, account for nearly all of the test-AUC gain (\S\ref{sec:exp}).}
  \Description{Block diagram of the solution pipeline. Non-sequential fields, dense features, and behavior
  sequences pass through their tokenizers into a shared token space, flow through a merged single-stream
  backbone of four attention blocks with a training-only masked-action auxiliary head, and end in a readout
  head and MLP classifier that produce the CVR logit.}
  \label{fig:arch}
\end{figure*}

Our model tokenizes every feature (sequential behavior and multi-field non-sequential attributes)
into a shared token space and reads it out into a CVR logit. We build on the official
PCVRHyFormer baseline, a latent-query transformer in the HyFormer and RankMixer line~\cite{hyformer,rankmixer}:
non-sequential fields form non-sequential (NS) tokens, user behavior forms sequence tokens, a stack of
attention blocks mixes them, and a readout head pools the tokens into a CVR logit. Figure~\ref{fig:arch}
diagrams the pipeline end to end: three input modalities tokenize into the shared space, a merged
single-stream backbone mixes them, and the dense-representation stack that feeds it is the largest single
contributor to test AUC (\S\ref{sec:exp}). We change one mechanism at a time
from this baseline, and the mechanisms below are grouped by the axis each belongs to.

\subsection{Dense-feature representation}
Dense features carry most of the exploitable signal, so we spend the most representational capacity on them.
The baseline pools the user dense fields into a few tokens and leaves heavy-tailed columns (magnitudes reach
$10^9$) untransformed. We instead split the pooled user dense fields into five subgroup tokens, apply
$\log(1+x)$ to the heavy-tailed columns, and add aligned-pair projections, item-side dense splitting, and
raw-statistic channels. This dense-representation stack is the single largest contributor in our ablation
(\S\ref{sec:exp}); Appendix~\ref{app:repro} gives the column-level recipe.

\subsection{Merged single-stream sequence modeling}
The baseline runs one sequence encoder per behavior domain, which stops events in different domains from
attending to each other in time order. We merge all domains into one timestamp-ordered stream and encode it
with stacked blocks that compress the sequence through decreasing top-$k$ retention, then read it out with
global NS tokens; this backbone follows the long-sequence line of LONGER~\cite{longer}. The merged stream
interleaves exposure, click, and conversion events of opposite polarity, and the shared embedding space hides
which events are negative. We add a target-conditioned polarity channel ported from TAPF~\cite{tapf}, validated on the
previous edition of this competition's dataset (TAAC-2025): a zero-initialized polarity bias per action-type field, plus a gate
$g_k=\sigma(W_g[h_k \,\|\, e_k \,\|\, b_{a_k}])$ that fuses each event with the candidate item before the
backbone. We also train the merged stream with a masked action modeling (MAM) auxiliary head~\cite{unipinrec}:
during training we mask each event's low-cardinality categorical attribute with probability 0.2 and predict
it with per-domain multi-class heads. This supplies dense per-position gradients and a denoising regularizer
that forces the encoder to rely on item content when the attribute is hidden. We port only this masked-action
objective, not UniPinRec's next-item retrieval loss, which would predict high-cardinality item IDs and invite
the memorization failure of \S\ref{sec:exp}.

\subsection{Optimization and regularization}
The optimizer is the second largest contributor. We optimize dense parameters with AMUSE~\cite{amuse}, a
Muon-family~\cite{muon} optimizer that orthogonalizes updates, and keep an exponential moving average of the
dense weights for best-model selection. Sparse embeddings use AdagradAR~\cite{adagradar}, Adagrad combined
with an adaptive regularizer that penalizes memorizing rare IDs. This regularizer is the change the non-temporal split most misvalues: it
lowers iid validation AUC yet raises the held-out leaderboard (\S\ref{sec:exp}). Appendix~\ref{app:repro}
lists learning rates, schedules, and seeds.

\section{Scaling}
\label{sec:scaling}
We sweep model width at two settings, $d_\text{model}\in\{160,320\}$. Widening $d_\text{model}$ from 160 to
320 adds 0.000104 test AUC, and halving it back costs 0.000319 in the ablation (Tables~\ref{tab:chain} and
\ref{tab:loo}), a positive but within-band effect.

Width was not the only capacity axis. The baseline also caps the query-token readout, the per-field
sub-dimension, and the number of retained NS tokens. These are capacity bottlenecks that no earlier step touched.
Extending the sequence budget, per domain (512${\to}$1024) or on the merged stream (1536${\to}$2432), did
not improve validation AUC in our runs, so we did not adopt it.

Capacity scaling along width produced small, within-band gains, and sequence-length
scaling produced none. The large test-AUC movements came from dense representation and optimization
instead (\S\ref{sec:exp}).

\section{Experiments}
\label{sec:exp}

\subsection{Setup}
\textbf{Dataset.} We use the Round-2 UniRec CVR dataset~\cite{unirecchallenge2026}: 34.82M click records
across 57,691 row groups. The label marks whether a click converts. Training and validation split by
row-group order rather than by time, so the iid validation set shares the training time window while the
leaderboard test set is a separate held-out dump.

\textbf{Metric.} We report AUC. Validation AUC is measured on the iid split; test AUC is the held-out
leaderboard score and is our verdict metric. The leaderboard keeps the maximum over a team's submissions.

\textbf{Protocol.} We train for five epochs on seven GPUs with six-way data parallelism, a 6{,}000-step
warmup into a constant learning rate (AMUSE is schedule free), and EMA-based best-model selection. The full
model has 530M parameters (179M dense, 351M sparse embeddings). Dense parameters use an orthogonalized
optimizer (AMUSE~\cite{amuse}); sparse embeddings use Adagrad with an adaptive regularizer. We treat
$\pm$0.0004 as the seed band: a seed replication of one mid-chain configuration moved test AUC by 0.00044,
and no per-step multi-seed estimate exists, so the band is a heuristic noise floor rather than a confidence
interval. We report absolute deltas and flag which fall inside it. Appendix~\ref{app:repro} lists hyperparameters, seeds, and the checkpoint protocol.

\subsection{Main Results}
The upgrade chain adds 0.014579 test AUC over the official baseline, and one dense step supplies more than
half of it. Table~\ref{tab:chain} traces fifteen single-variable steps from 0.813237 to 0.827816. The chain
is greedy forward selection against the leaderboard, and each delta is one run measured against a
single-replication noise floor, so steps below the floor read as adopted, not as attributed mechanism
contributions; attribution comes from the leave-one-out ablation below. Splitting
the user dense fields into subgroup tokens alone adds 0.007988, or 55\% of the total gain. The four
dense-representation steps together add 0.011183, while no step that touches the sequence encoder or its
training objective (LONGER, MAM, merged backbone, TAPF) adds more than 0.0003. Our final leaderboard
submission adds a per-token FFN and full-data training without the validation holdout, submitting per-epoch
snapshots under the leaderboard's max rule; it reaches 0.828535 and places 10th.

\begin{table}[t]
\centering
\caption{The single-variable upgrade chain. Each step changes one mechanism over the previous row. Test AUC
is the held-out leaderboard score; $\Delta$ is the gain over the previous adopted step. Deltas are single-run
(round-2 has no per-step seed band).}
\label{tab:chain}
\small
\begin{tabular}{@{}llrr@{}}
\toprule
Step (single-variable change) & Type & Test AUC & $\Delta$AUC \\
\midrule
Official baseline (PCVRHyFormer) & n/a & 0.813237 & n/a \\
$+$ AMUSE optimizer~\cite{amuse} & optim. & 0.813629 & $+$0.000392 \\
$+$ per-domain time embedding & struct. & 0.814387 & $+$0.000758 \\
$+$ user dense field split & dense & 0.822375 & $+$0.007988 \\
$+$ item dense split, log1p\textsuperscript{a} & dense & 0.823831 & $+$0.001456 \\
$+$ user raw-stat log1p & dense & 0.824995 & $+$0.001164 \\
$+$ dense-weight EMA & train & 0.825472 & $+$0.000477 \\
$+$ width ($d_\text{model}{=}160$) & capacity & 0.825837 & $+$0.000365 \\
$+$ aligned-pair dense & dense & 0.826412 & $+$0.000575 \\
$+$ LONGER sequence encoder~\cite{longer} & seq. & 0.826477 & $+$0.000065 \\
$+$ NS semantic grouping & token. & 0.826596 & $+$0.000119 \\
$+$ AdagradAR sparse reg. & optim. & 0.827227 & $+$0.000631 \\
$+$ MAM auxiliary head & seq. & 0.827380 & $+$0.000153 \\
$+$ merged single-stream backbone & seq. & 0.827611 & $+$0.000231 \\
$+$ TAPF polarity channel~\cite{tapf} & seq. & 0.827712 & $+$0.000101 \\
$+$ width ($d_\text{model}{=}320$) & capacity & 0.827816 & $+$0.000104 \\
\midrule
Final submission (multi-variable) & n/a & \textbf{0.828535} & n/a \\
\bottomrule
\end{tabular}
\\[3pt]
{\footnotesize \textsuperscript{a}Bundles an aligned-pair log1p step with no separately submitted score.}
\end{table}

Dense-feature representation, not sequence modeling, drives the chain: no sequence-side step clears the
seed band.

\subsection{Ablation Study}
Leave-one-out ablation attributes nearly all of the gain to the dense-representation stack and the optimizer. The full
model is the chain endpoint plus a per-token FFN (the final submission's architecture), retrained with the
validation split restored so that every ablation shares one protocol; it scores 0.827932, and the remaining
0.000603 to the final submission comes from full-data training and snapshot selection.
Table~\ref{tab:loo} removes one component at a time from this model. Removing the dense-representation
stack drops test AUC by 0.009496, an order of magnitude larger than any other component. Replacing the
orthogonalized optimizer with AdamW~\cite{adamw} drops 0.002821, the second largest. The AdagradAR sparse
regularizer accounts for 0.000890. The remaining seven components each fall within or adjacent to the
$\pm$0.0004 seed band. The merged-backbone row is a joint removal (the model gates TAPF and the per-token
FFN to the merged stream); subtracting the TAPF and FFN rows under an additivity assumption leaves the
backbone itself at $+$0.00007, indistinguishable from zero, though a clean backbone-only run does not exist.

\begin{table}[t]
\centering
\caption{Leave-one-out ablation from the full model. Each row removes one component and reports the
test-AUC drop. The dense-representation stack and the orthogonalized optimizer dominate; the rest fall within
the $\pm$0.0004 seed band.}
\label{tab:loo}
\small
\begin{tabular}{@{}lrr@{}}
\toprule
Removed component & Test AUC & $\Delta$AUC \\
\midrule
Full model & 0.827932 & n/a \\
\midrule
$-$ dense-representation stack & 0.818436 & $-$0.009496 \\
$-$ AMUSE optimizer ($\to$ AdamW) & 0.825111 & $-$0.002821 \\
$-$ AdagradAR sparse reg. & 0.827042 & $-$0.000890 \\
$-$ merged single-stream backbone\textsuperscript{a} & 0.827467 & $-$0.000465 \\
$-$ TAPF polarity channel & 0.827538 & $-$0.000394 \\
$-$ dense-weight EMA & 0.827581 & $-$0.000351 \\
$-$ width ($d_\text{model}\,320{\to}160$) & 0.827613 & $-$0.000319 \\
$-$ per-domain time embedding & 0.827717 & $-$0.000215 \\
$-$ per-token FFN & 0.827790 & $-$0.000142 \\
$-$ MAM auxiliary head & 0.827928 & $-$0.000004 \\
\bottomrule
\end{tabular}
\\[3pt]
{\footnotesize \textsuperscript{a}Removing the merged backbone also disables TAPF and per-token FFN, which the
model gates to it.}
\end{table}

Leave-one-out attribution corrects the chain's ordering bias, which runs both ways. The optimizer entered
the chain early, on a weak parent, and added only 0.000392 there (Table~\ref{tab:chain}); removing it from
the final model costs 0.002821, seven times larger. The per-domain time embedding shows the reverse: it
added 0.000758 in the chain but costs only 0.000215 when removed. Chain deltas mismeasure components by
entry position, so we attribute contribution by leave-one-out from the final model. The chain's headline
survives this correction: removing the dense stack costs 0.009496, or 65\% of the total gain.

Two variants match or beat the full model, and we report both. Replacing AMUSE with Muon~\cite{muon} raises
test by 0.000273, so the optimizer axis reads AdamW ($0.825111$) $<$ AMUSE ($0.827932$) $\approx$ Muon
($0.828205$), with AMUSE and Muon indistinguishable within the seed band. Removing MAM
changes test by 0.000004, so it contributes nothing to the final model.

The gains concentrate in dense representation and the optimizer; sequence-modeling
refinements sit individually below our single-estimate noise floor. This matches the user-behavior feature
probes that produced no test gain (\S\ref{sec:discussion}): at this scale and split, test AUC moves through
dense features and optimization.

\subsection{Leaderboard Insights}
Under the non-temporal split, validation AUC is an inflated and sometimes sign-flipped proxy for the
leaderboard. The full model scores 0.842 on iid validation but 0.828 on test, a 0.014 gap. Because training
and validation share one time window while the test dump does not, the split rewards memorizing that window.

Regularization and capacity changes expose the sign flip. Adding the AdagradAR sparse regularizer lowers
validation AUC by 0.00031 yet raises test AUC by 0.000631. An anti-memorization change looks harmful on
iid validation but helps on the held-out dump. Enlarging high-cardinality ID embeddings shows the mirror
image in four separate runs from the competition's earlier round: validation rises while test falls, as
memorized IDs pay off in-window and misfire across dumps.

The leaderboard keeps the maximum over submissions, which makes any positive-expectation change worth
submitting. A submission can only raise the displayed score, never lower it, so selective aggression
dominates preservation once validation is an unreliable filter.

Under this competition's non-temporal split, validation AUC is a biased proxy, and it inverts the sign of
the regularization and capacity changes that help generalization. An out-of-time proxy that re-splits the
training dump by timestamp does not recover the AdagradAR sign either ($-$0.0006 out-of-time vs.\ $+$0.0006
on test), so for these changes the divergence reflects dump-to-dump distribution shift, not just time
ordering inside one dump. Verdicts must come from the held-out leaderboard, not from in-dump validation.

\section{Discussion: Limitations and Future Work}
\label{sec:discussion}
\textbf{Negative results.} Several mechanisms that help in the literature did not survive here, at two
evidence levels. Three were screened out on validation: an explicit DCN-v2 feature-crossing
bypass~\cite{dcnv2} ($-$0.00121), a target-conditioned query gate~\cite{lens} ($-$0.00258), and a
category-level target-attention~\cite{din} retrieval over the user's category history~\cite{uxsid}
($-$0.00382). Validation is a biased filter (\S\ref{sec:exp}), but every sign inversion we observed came
with a validation drop under 0.0004, an order of magnitude smaller than these regressions, so we judged a
leaderboard submission on them negative expected value under the three-per-day budget. Two were measured on
test: a gated self-attention sequence encoder gained 0.00037, within the seed band, at 50\% more training
time, so we did not adopt it; and hand-built user-behavior features (time-of-day, intent, price ordinals)
produced no test gain over the dense-and-optimizer stack.

\textbf{Limitations and future work.} Our verdicts depend on a held-out leaderboard capped at three
submissions per day, and our per-step deltas are single-run, so we report absolute deltas without a seed band
on the chain. The non-temporal split makes validation a biased filter; the out-of-time proxy we built recovers an honest
scale (its AUC sits near test rather than near iid validation) but still misreads the anti-memorization axis
(\S\ref{sec:exp}), so a time-ordered evaluation across dumps remains future work. Multi-seed estimates would
sharpen every within-band claim in \S\ref{sec:exp}.

\section{Conclusion}
\label{sec:conclusion}
A 15-step single-variable chain and a leave-one-out ablation locate where industrial CVR AUC comes from:
dense-feature representation and optimizer choice supply nearly all of a 0.014579 gain, while sequence-modeling
refinements fall within the seed band. Under the competition's non-temporal split, validation AUC inflates and sometimes
inverts against the held-out leaderboard, so verdicts must come from the leaderboard. Our final model reaches
0.828535 test AUC and 10th place in the KDD Cup 2026 UniRec Challenge.

\bibliographystyle{ACM-Reference-Format}
\bibliography{references}


\begin{thebibliography}{23}


\ifx \showCODEN    \undefined \def \showCODEN     #1{\unskip}     \fi
\ifx \showISBNx    \undefined \def \showISBNx     #1{\unskip}     \fi
\ifx \showISBNxiii \undefined \def \showISBNxiii  #1{\unskip}     \fi
\ifx \showISSN     \undefined \def \showISSN      #1{\unskip}     \fi
\ifx \showLCCN     \undefined \def \showLCCN      #1{\unskip}     \fi
\ifx \shownote     \undefined \def \shownote      #1{#1}          \fi
\ifx \showarticletitle \undefined \def \showarticletitle #1{#1}   \fi
\ifx \showURL      \undefined \def \showURL       {\relax}        \fi
\providecommand\bibfield[2]{#2}
\providecommand\bibinfo[2]{#2}
\providecommand\natexlab[1]{#1}
\providecommand\showeprint[2][]{arXiv:#2}

\bibitem[Chai et~al\mbox{.}(2025)]%
        {longer}
\bibfield{author}{\bibinfo{person}{Zheng Chai}, \bibinfo{person}{Qin Ren}, \bibinfo{person}{Xijun Xiao}, \bibinfo{person}{Huizhi Yang}, \bibinfo{person}{Bo Han}, \bibinfo{person}{Sijun Zhang}, \bibinfo{person}{Di Chen}, \bibinfo{person}{Hui Lu}, \bibinfo{person}{Wenlin Zhao}, \bibinfo{person}{Lele Yu}, \bibinfo{person}{Xionghang Xie}, \bibinfo{person}{Shiru Ren}, \bibinfo{person}{Xiang Sun}, \bibinfo{person}{Yaocheng Tan}, \bibinfo{person}{Peng Xu}, \bibinfo{person}{Yuchao Zheng}, {and} \bibinfo{person}{Di Wu}.} \bibinfo{year}{2025}\natexlab{}.
\newblock \showarticletitle{{LONGER: Scaling Up Long Sequence Modeling in Industrial Recommenders}}. In \bibinfo{booktitle}{\emph{Proceedings of the 19th ACM Conference on Recommender Systems (RecSys '25)}}. \bibinfo{publisher}{Association for Computing Machinery}, \bibinfo{address}{New York, NY, USA}, \bibinfo{pages}{247--256}.
\newblock
\showeprint[arxiv]{2505.04421}~[cs.IR]
\href{https://doi.org/10.1145/3705328.3748065}{doi:\nolinkurl{10.1145/3705328.3748065}}


\bibitem[Chapelle(2014)]%
        {chapelle2014delayed}
\bibfield{author}{\bibinfo{person}{Olivier Chapelle}.} \bibinfo{year}{2014}\natexlab{}.
\newblock \showarticletitle{Modeling Delayed Feedback in Display Advertising}. In \bibinfo{booktitle}{\emph{Proceedings of the 20th ACM SIGKDD International Conference on Knowledge Discovery and Data Mining (KDD '14)}}. \bibinfo{publisher}{Association for Computing Machinery}, \bibinfo{address}{New York, NY, USA}, \bibinfo{pages}{1097--1105}.
\newblock
\href{https://doi.org/10.1145/2623330.2623634}{doi:\nolinkurl{10.1145/2623330.2623634}}


\bibitem[Cheng et~al\mbox{.}(2026)]%
        {tapf}
\bibfield{author}{\bibinfo{person}{Zexuan Cheng}, \bibinfo{person}{Yue Liu}, \bibinfo{person}{Jun Zhang}, {and} \bibinfo{person}{Jie Jiang}.} \bibinfo{year}{2026}\natexlab{}.
\newblock \bibinfo{title}{{Beyond Positive Signals: Unlocking Implicit Negative Behaviors for Enhanced Sequential User Modeling}}.
\newblock \bibinfo{howpublished}{arXiv preprint arXiv:2606.15252}.
\newblock
\showeprint[arxiv]{2606.15252}~[cs.IR]


\bibitem[Hou et~al\mbox{.}(2026)]%
        {kunlun}
\bibfield{author}{\bibinfo{person}{Bojian Hou}, \bibinfo{person}{Xiaolong Liu}, \bibinfo{person}{Xiaoyi Liu}, \bibinfo{person}{Jiaqi Xu}, \bibinfo{person}{Yasmine Badr}, \bibinfo{person}{Mengyue Hang}, \bibinfo{person}{Sudhanshu Chanpuriya}, \bibinfo{person}{Junqing Zhou}, \bibinfo{person}{Yuhang Yang}, \bibinfo{person}{Han Xu}, \bibinfo{person}{Qiuling Suo}, \bibinfo{person}{Laming Chen}, \bibinfo{person}{Yuxi Hu}, \bibinfo{person}{Jiasheng Zhang}, \bibinfo{person}{Huaqing Xiong}, \bibinfo{person}{Yuzhen Huang}, \bibinfo{person}{Chao Chen}, \bibinfo{person}{Yue Dong}, \bibinfo{person}{Yi Yang}, \bibinfo{person}{Shuo Chang}, \bibinfo{person}{Xiaorui Gan}, \bibinfo{person}{Wenlin Chen}, \bibinfo{person}{Santanu Kolay}, \bibinfo{person}{Darren Liu}, \bibinfo{person}{Jade Nie}, \bibinfo{person}{Chunzhi Yang}, \bibinfo{person}{Ellie Wen}, \bibinfo{person}{Jiyan Yang}, {and} \bibinfo{person}{Huayu Li}.} \bibinfo{year}{2026}\natexlab{}.
\newblock \bibinfo{title}{{Kunlun: Establishing Scaling Laws for Massive-Scale Recommendation Systems through Unified Architecture Design}}.
\newblock \bibinfo{howpublished}{arXiv preprint arXiv:2602.10016}.
\newblock
\showeprint[arxiv]{2602.10016}~[cs.IR]


\bibitem[Huang et~al\mbox{.}(2026b)]%
        {mixformer}
\bibfield{author}{\bibinfo{person}{Xu Huang}, \bibinfo{person}{Hao Zhang}, \bibinfo{person}{Zhifang Fan}, \bibinfo{person}{Yunwen Huang}, \bibinfo{person}{Zhuoxing Wei}, \bibinfo{person}{Zheng Chai}, \bibinfo{person}{Jinan Ni}, \bibinfo{person}{Yuchao Zheng}, {and} \bibinfo{person}{Qiwei Chen}.} \bibinfo{year}{2026}\natexlab{b}.
\newblock \bibinfo{title}{{MixFormer: Co-Scaling Up Dense and Sequence in Industrial Recommenders}}.
\newblock \bibinfo{howpublished}{arXiv preprint arXiv:2602.14110}.
\newblock
\showeprint[arxiv]{2602.14110}~[cs.IR]


\bibitem[Huang et~al\mbox{.}(2026a)]%
        {hyformer}
\bibfield{author}{\bibinfo{person}{Yunwen Huang}, \bibinfo{person}{Shiyong Hong}, \bibinfo{person}{Xijun Xiao}, \bibinfo{person}{Jinqiu Jin}, \bibinfo{person}{Xuanyuan Luo}, \bibinfo{person}{Zhe Wang}, \bibinfo{person}{Zheng Chai}, \bibinfo{person}{Shikang Wu}, \bibinfo{person}{Yuchao Zheng}, {and} \bibinfo{person}{Jingjian Lin}.} \bibinfo{year}{2026}\natexlab{a}.
\newblock \bibinfo{title}{{HyFormer: Revisiting the Roles of Sequence Modeling and Feature Interaction in CTR Prediction}}.
\newblock \bibinfo{howpublished}{arXiv preprint arXiv:2601.12681}.
\newblock
\showeprint[arxiv]{2601.12681}~[cs.IR]


\bibitem[Ji et~al\mbox{.}(2023)]%
        {ji2023leakage}
\bibfield{author}{\bibinfo{person}{Yitong Ji}, \bibinfo{person}{Aixin Sun}, \bibinfo{person}{Jie Zhang}, {and} \bibinfo{person}{Chenliang Li}.} \bibinfo{year}{2023}\natexlab{}.
\newblock \showarticletitle{A Critical Study on Data Leakage in Recommender System Offline Evaluation}.
\newblock \bibinfo{journal}{\emph{ACM Transactions on Information Systems}} \bibinfo{volume}{41}, \bibinfo{number}{3}, Article \bibinfo{articleno}{75} (\bibinfo{year}{2023}), \bibinfo{numpages}{27}~pages.
\newblock
\href{https://doi.org/10.1145/3569930}{doi:\nolinkurl{10.1145/3569930}}


\bibitem[Jordan et~al\mbox{.}(2024)]%
        {muon}
\bibfield{author}{\bibinfo{person}{Keller Jordan}, \bibinfo{person}{Yuchen Jin}, \bibinfo{person}{Vlado Boza}, \bibinfo{person}{Jiacheng You}, \bibinfo{person}{Franz Cesista}, \bibinfo{person}{Laker Newhouse}, {and} \bibinfo{person}{Jeremy Bernstein}.} \bibinfo{year}{2024}\natexlab{}.
\newblock \bibinfo{title}{{Muon: An optimizer for hidden layers in neural networks}}.
\newblock \bibinfo{howpublished}{\url{https://kellerjordan.github.io/posts/muon/}}.
\newblock
\shownote{Blog post, 8 December 2024. Code: \url{https://github.com/KellerJordan/Muon}}.
\newblock


\bibitem[{KDD Cup 2026 Tencent UniRec Challenge Organizing Committee}(2026)]%
        {unirecchallenge2026}
\bibfield{author}{\bibinfo{person}{{KDD Cup 2026 Tencent UniRec Challenge Organizing Committee}}.} \bibinfo{year}{2026}\natexlab{}.
\newblock \bibinfo{title}{{KDD Cup 2026 Tencent UniRec Challenge}}.
\newblock \bibinfo{howpublished}{\url{https://algo.qq.com/}}.
\newblock


\bibitem[Kim et~al\mbox{.}(2026)]%
        {amuse}
\bibfield{author}{\bibinfo{person}{Jueun Kim}, \bibinfo{person}{Baekrok Shin}, \bibinfo{person}{Jihun Yun}, \bibinfo{person}{Beomhan Baek}, \bibinfo{person}{Minhak Song}, {and} \bibinfo{person}{Chulhee Yun}.} \bibinfo{year}{2026}\natexlab{}.
\newblock \bibinfo{title}{{AMUSE: Anytime Muon with Stable Gradient Evaluation}}.
\newblock \bibinfo{howpublished}{arXiv preprint arXiv:2605.22432}.
\newblock
\showeprint[arxiv]{2605.22432}~[cs.LG]


\bibitem[Li et~al\mbox{.}(2026)]%
        {unipinrec}
\bibfield{author}{\bibinfo{person}{Hanyu Li}, \bibinfo{person}{Yi-Ping Hsu}, \bibinfo{person}{Aditya Mantha}, \bibinfo{person}{Prabhat Agarwal}, \bibinfo{person}{Laksh Bhasin}, \bibinfo{person}{Jialu Wang}, \bibinfo{person}{Hongtao Lin}, \bibinfo{person}{Bella Huang}, \bibinfo{person}{Yaxin Li}, \bibinfo{person}{Xinyi Li}, \bibinfo{person}{Chuxi Wang}, \bibinfo{person}{Kousik Rajesh}, \bibinfo{person}{Hooshmand Shokri~Razaghi}, \bibinfo{person}{Shunyao Li}, \bibinfo{person}{Zongyue Qin}, \bibinfo{person}{Jaewon Yang}, \bibinfo{person}{James Li}, \bibinfo{person}{Dhruvil~Deven Badani}, \bibinfo{person}{Jiajing Xu}, {and} \bibinfo{person}{Charles Rosenberg}.} \bibinfo{year}{2026}\natexlab{}.
\newblock \bibinfo{title}{{UniPinRec: Unifying Generative Retrieval and Ranking at Pinterest Scale}}.
\newblock \bibinfo{howpublished}{arXiv preprint arXiv:2606.00422}.
\newblock
\showeprint[arxiv]{2606.00422}~[cs.IR]


\bibitem[Li and Lyu(2025)]%
        {adagradar}
\bibfield{author}{\bibinfo{person}{Mang Li} {and} \bibinfo{person}{Wei Lyu}.} \bibinfo{year}{2025}\natexlab{}.
\newblock \bibinfo{title}{{Adaptive Regularization for Large-Scale Sparse Feature Embedding Models}}.
\newblock \bibinfo{howpublished}{arXiv preprint arXiv:2511.06374}.
\newblock
\showeprint[arxiv]{2511.06374}~[cs.LG]


\bibitem[Loshchilov and Hutter(2019)]%
        {adamw}
\bibfield{author}{\bibinfo{person}{Ilya Loshchilov} {and} \bibinfo{person}{Frank Hutter}.} \bibinfo{year}{2019}\natexlab{}.
\newblock \showarticletitle{{Decoupled Weight Decay Regularization}}. In \bibinfo{booktitle}{\emph{7th International Conference on Learning Representations (ICLR 2019)}}. \bibinfo{publisher}{OpenReview.net}.
\newblock
\showeprint[arxiv]{1711.05101}~[cs.LG]


\bibitem[Meng et~al\mbox{.}(2020)]%
        {meng2020splitting}
\bibfield{author}{\bibinfo{person}{Zaiqiao Meng}, \bibinfo{person}{Richard McCreadie}, \bibinfo{person}{Craig Macdonald}, {and} \bibinfo{person}{Iadh Ounis}.} \bibinfo{year}{2020}\natexlab{}.
\newblock \showarticletitle{Exploring Data Splitting Strategies for the Evaluation of Recommendation Models}. In \bibinfo{booktitle}{\emph{Proceedings of the 14th ACM Conference on Recommender Systems (RecSys '20)}}. \bibinfo{publisher}{Association for Computing Machinery}, \bibinfo{address}{New York, NY, USA}, \bibinfo{pages}{681--686}.
\newblock
\href{https://doi.org/10.1145/3383313.3418479}{doi:\nolinkurl{10.1145/3383313.3418479}}


\bibitem[Wang et~al\mbox{.}(2021)]%
        {dcnv2}
\bibfield{author}{\bibinfo{person}{Ruoxi Wang}, \bibinfo{person}{Rakesh Shivanna}, \bibinfo{person}{Derek~Z. Cheng}, \bibinfo{person}{Sagar Jain}, \bibinfo{person}{Dong Lin}, \bibinfo{person}{Lichan Hong}, {and} \bibinfo{person}{Ed~H. Chi}.} \bibinfo{year}{2021}\natexlab{}.
\newblock \showarticletitle{{DCN V2: Improved Deep \& Cross Network and Practical Lessons for Web-scale Learning to Rank Systems}}. In \bibinfo{booktitle}{\emph{Proceedings of the Web Conference 2021 (WWW '21)}}. \bibinfo{publisher}{Association for Computing Machinery}, \bibinfo{address}{New York, NY, USA}, \bibinfo{pages}{1785--1797}.
\newblock
\showeprint[arxiv]{2008.13535}~[cs.IR]
\href{https://doi.org/10.1145/3442381.3450078}{doi:\nolinkurl{10.1145/3442381.3450078}}


\bibitem[Wang et~al\mbox{.}(2026)]%
        {lens}
\bibfield{author}{\bibinfo{person}{Yuan Wang}, \bibinfo{person}{Yue Liu}, \bibinfo{person}{Jun Zhang}, {and} \bibinfo{person}{Jie Jiang}.} \bibinfo{year}{2026}\natexlab{}.
\newblock \bibinfo{title}{{LENS: A Staged Design for Interaction Granularity in Sequential CTR Prediction}}.
\newblock \bibinfo{howpublished}{arXiv preprint arXiv:2605.25583}.
\newblock
\showeprint[arxiv]{2605.25583}~[cs.IR]


\bibitem[Zeng et~al\mbox{.}(2025)]%
        {interformer}
\bibfield{author}{\bibinfo{person}{Zhichen Zeng}, \bibinfo{person}{Xiaolong Liu}, \bibinfo{person}{Mengyue Hang}, \bibinfo{person}{Xiaoyi Liu}, \bibinfo{person}{Qinghai Zhou}, \bibinfo{person}{Chaofei Yang}, \bibinfo{person}{Yiqun Liu}, \bibinfo{person}{Yichen Ruan}, \bibinfo{person}{Laming Chen}, \bibinfo{person}{Yuxin Chen}, \bibinfo{person}{Yujia Hao}, \bibinfo{person}{Jiaqi Xu}, \bibinfo{person}{Jade Nie}, \bibinfo{person}{Xi Liu}, \bibinfo{person}{Buyun Zhang}, \bibinfo{person}{Wei Wen}, \bibinfo{person}{Siyang Yuan}, \bibinfo{person}{Hang Yin}, \bibinfo{person}{Xin Zhang}, \bibinfo{person}{Kai Wang}, \bibinfo{person}{Wen-Yen Chen}, \bibinfo{person}{Yiping Han}, \bibinfo{person}{Huayu Li}, \bibinfo{person}{Chunzhi Yang}, \bibinfo{person}{Bo Long}, \bibinfo{person}{Philip~S. Yu}, \bibinfo{person}{Hanghang Tong}, {and} \bibinfo{person}{Jiyan Yang}.} \bibinfo{year}{2025}\natexlab{}.
\newblock \showarticletitle{{InterFormer: Effective Heterogeneous Interaction Learning for Click-Through Rate Prediction}}. In \bibinfo{booktitle}{\emph{Proceedings of the 34th ACM International Conference on Information and Knowledge Management (CIKM '25)}}. \bibinfo{publisher}{Association for Computing Machinery}, \bibinfo{address}{New York, NY, USA}, \bibinfo{pages}{6225--6233}.
\newblock
\showeprint[arxiv]{2411.09852}~[cs.IR]
\href{https://doi.org/10.1145/3746252.3761527}{doi:\nolinkurl{10.1145/3746252.3761527}}


\bibitem[Zhai et~al\mbox{.}(2024)]%
        {hstu}
\bibfield{author}{\bibinfo{person}{Jiaqi Zhai}, \bibinfo{person}{Lucy Liao}, \bibinfo{person}{Xing Liu}, \bibinfo{person}{Yueming Wang}, \bibinfo{person}{Rui Li}, \bibinfo{person}{Xuan Cao}, \bibinfo{person}{Leon Gao}, \bibinfo{person}{Zhaojie Gong}, \bibinfo{person}{Fangda Gu}, \bibinfo{person}{Michael He}, \bibinfo{person}{Yinghai Lu}, {and} \bibinfo{person}{Yu Shi}.} \bibinfo{year}{2024}\natexlab{}.
\newblock \showarticletitle{{Actions Speak Louder than Words: Trillion-Parameter Sequential Transducers for Generative Recommendations}}. In \bibinfo{booktitle}{\emph{Proceedings of the 41st International Conference on Machine Learning (ICML 2024)}}. \bibinfo{publisher}{PMLR}, \bibinfo{pages}{58484--58509}.
\newblock
\showeprint[arxiv]{2402.17152}~[cs.LG]


\bibitem[Zhang et~al\mbox{.}(2024)]%
        {wukong}
\bibfield{author}{\bibinfo{person}{Buyun Zhang}, \bibinfo{person}{Liang Luo}, \bibinfo{person}{Yuxin Chen}, \bibinfo{person}{Jade Nie}, \bibinfo{person}{Xi Liu}, \bibinfo{person}{Daifeng Guo}, \bibinfo{person}{Yanli Zhao}, \bibinfo{person}{Shen Li}, \bibinfo{person}{Yuchen Hao}, \bibinfo{person}{Yantao Yao}, \bibinfo{person}{Guna Lakshminarayanan}, \bibinfo{person}{Ellie~Dingqiao Wen}, \bibinfo{person}{Jongsoo Park}, \bibinfo{person}{Maxim Naumov}, {and} \bibinfo{person}{Wenlin Chen}.} \bibinfo{year}{2024}\natexlab{}.
\newblock \showarticletitle{{Wukong: Towards a Scaling Law for Large-Scale Recommendation}}. In \bibinfo{booktitle}{\emph{Proceedings of the 41st International Conference on Machine Learning (ICML 2024)}}. \bibinfo{publisher}{PMLR}, \bibinfo{pages}{59421--59434}.
\newblock
\showeprint[arxiv]{2403.02545}~[cs.LG]


\bibitem[Zhang et~al\mbox{.}(2026b)]%
        {uxsid}
\bibfield{author}{\bibinfo{person}{Hongwei Zhang}, \bibinfo{person}{Qiqiang Zhong}, \bibinfo{person}{Jiangxia Cao}, \bibinfo{person}{Yiyang Lv}, \bibinfo{person}{Huanjie Wang}, \bibinfo{person}{Liwei Guan}, \bibinfo{person}{Jing Yao}, \bibinfo{person}{Yiyu Wang}, \bibinfo{person}{Junfeng Shu}, \bibinfo{person}{Zhaojie Liu}, {and} \bibinfo{person}{Han Li}.} \bibinfo{year}{2026}\natexlab{b}.
\newblock \bibinfo{title}{{UxSID: Semantic-Aware User Interests Modeling for Ultra-Long Sequence}}.
\newblock \bibinfo{howpublished}{arXiv preprint arXiv:2605.09040}.
\newblock
\showeprint[arxiv]{2605.09040}~[cs.AI]


\bibitem[Zhang et~al\mbox{.}(2026a)]%
        {onetrans}
\bibfield{author}{\bibinfo{person}{Zhaoqi Zhang}, \bibinfo{person}{Haolei Pei}, \bibinfo{person}{Jun Guo}, \bibinfo{person}{Tianyu Wang}, \bibinfo{person}{Yufei Feng}, \bibinfo{person}{Hui Sun}, \bibinfo{person}{Shaowei Liu}, {and} \bibinfo{person}{Aixin Sun}.} \bibinfo{year}{2026}\natexlab{a}.
\newblock \showarticletitle{{OneTrans: Unified Feature Interaction and Sequence Modeling with One Transformer in Industrial Recommender}}. In \bibinfo{booktitle}{\emph{Proceedings of the ACM Web Conference 2026 (WWW '26)}}. \bibinfo{publisher}{Association for Computing Machinery}, \bibinfo{address}{New York, NY, USA}, \bibinfo{pages}{8162--8170}.
\newblock
\showeprint[arxiv]{2510.26104}~[cs.IR]
\href{https://doi.org/10.1145/3774904.3792838}{doi:\nolinkurl{10.1145/3774904.3792838}}


\bibitem[Zhou et~al\mbox{.}(2018)]%
        {din}
\bibfield{author}{\bibinfo{person}{Guorui Zhou}, \bibinfo{person}{Chengru Song}, \bibinfo{person}{Xiaoqiang Zhu}, \bibinfo{person}{Ying Fan}, \bibinfo{person}{Han Zhu}, \bibinfo{person}{Xiao Ma}, \bibinfo{person}{Yanghui Yan}, \bibinfo{person}{Junqi Jin}, \bibinfo{person}{Han Li}, {and} \bibinfo{person}{Kun Gai}.} \bibinfo{year}{2018}\natexlab{}.
\newblock \showarticletitle{{Deep Interest Network for Click-Through Rate Prediction}}. In \bibinfo{booktitle}{\emph{Proceedings of the 24th ACM SIGKDD International Conference on Knowledge Discovery \& Data Mining (KDD '18)}}. \bibinfo{publisher}{Association for Computing Machinery}, \bibinfo{address}{New York, NY, USA}, \bibinfo{pages}{1059--1068}.
\newblock
\showeprint[arxiv]{1706.06978}~[stat.ML]
\href{https://doi.org/10.1145/3219819.3219823}{doi:\nolinkurl{10.1145/3219819.3219823}}


\bibitem[Zhu et~al\mbox{.}(2025)]%
        {rankmixer}
\bibfield{author}{\bibinfo{person}{Jie Zhu}, \bibinfo{person}{Zhifang Fan}, \bibinfo{person}{Xiaoxie Zhu}, \bibinfo{person}{Yuchen Jiang}, \bibinfo{person}{Hangyu Wang}, \bibinfo{person}{Xintian Han}, \bibinfo{person}{Haoran Ding}, \bibinfo{person}{Xinmin Wang}, \bibinfo{person}{Wenlin Zhao}, \bibinfo{person}{Zhen Gong}, \bibinfo{person}{Huizhi Yang}, \bibinfo{person}{Zheng Chai}, \bibinfo{person}{Zhe Chen}, \bibinfo{person}{Yuchao Zheng}, \bibinfo{person}{Qiwei Chen}, \bibinfo{person}{Feng Zhang}, \bibinfo{person}{Xun Zhou}, \bibinfo{person}{Peng Xu}, \bibinfo{person}{Xiao Yang}, \bibinfo{person}{Di Wu}, {and} \bibinfo{person}{Zuotao Liu}.} \bibinfo{year}{2025}\natexlab{}.
\newblock \showarticletitle{{RankMixer: Scaling Up Ranking Models in Industrial Recommenders}}. In \bibinfo{booktitle}{\emph{Proceedings of the 34th ACM International Conference on Information and Knowledge Management (CIKM '25)}}. \bibinfo{publisher}{Association for Computing Machinery}, \bibinfo{address}{New York, NY, USA}, \bibinfo{pages}{6309--6316}.
\newblock
\showeprint[arxiv]{2507.15551}~[cs.IR]
\href{https://doi.org/10.1145/3746252.3761507}{doi:\nolinkurl{10.1145/3746252.3761507}}


\end{thebibliography}

\appendix

\section{Reproducibility}
\label{app:repro}
We train on seven GPUs (six-way DDP via \texttt{torchrun}) for five epochs at batch size 1024, with a
standard row-group split (the last 10\% of row groups as validation) and best-model selection by validation
AUC. All runs use a single fixed seed; we did not run multi-seed estimates (\S\ref{sec:discussion}).
Table~\ref{tab:config} lists the model and optimizer configuration.

\begin{table}[htbp]
\centering
\caption{Configuration of the full model.}
\label{tab:config}
\small
\begin{tabular}{@{}lp{5.3cm}@{}}
\toprule
Group & Setting \\
\midrule
Backbone & merged single-stream, 4 blocks, merged length 1536 \\
Width & $d_\text{model}{=}320$, embedding dim 128, 4 heads, FFN mult 8 \\
Readout & flatten NS tokens $\to$ linear; per-token FFN \\
NS tokens & group tokenizer; 4 user, 2 item NS tokens \\
Sequence & per-block top-$k$ retention $[512, 171, 57, 30]$ \\
Time & per-domain time-bucket embedding \\
Dense & field split, aligned-pair, item split, raw-statistic channels \\
Optimizer (dense) & AMUSE: lr 0.01, momentum 0.95, wd 0.05, warmup 6000 \\
Optimizer (sparse) & AdagradAR ($\alpha{=}10^{-4}$) \\
EMA & decay 0.999, start step 100 \\
Aux tasks & MAM (mask 0.2, weight 0.1, window 50); TAPF polarity (dim 32) \\
\bottomrule
\end{tabular}
\end{table}

\textbf{Dense-representation recipe.} Five flags implement the dense stack, and the ablation row
``$-$~dense-representation stack'' (Table~\ref{tab:loo}) removes exactly these five. The 17 user dense fields split
into five subgroup tokens, each projected by a linear layer with LayerNorm and SiLU: a pretrained sum
embedding (field 61), an LMF embedding (field 87), three unit-norm aligned statistics (fields 89--91), an
aligned-pair group (fields 62--66, 118, 121), and one token that averages a mid-scale head (fields 120, 130;
LayerNorm only) with a raw-statistic head (fields 123, 131, 132; $\log(1{+}\max(x,0))$ before projection).
The aligned-pair group does not project raw values: each dense column pairs with an integer column of tag
IDs, and its token pools the tag-embedding table shared with the categorical tokenizer, weighting each tag
by $\log(1{+}\max(x,0))$ of its count and normalizing by the weight sum. The four item dense fields split
the same way into a pretrained-vector head (field 127), a mid-scale head (field 128), and a log1p
raw-statistic head (fields 124, 129), averaged into one token. Without these flags, user and item dense
features each collapse to a single linear projection, and the heavy-tailed columns dominate the gradient
(\S\ref{sec:method}).

Checkpoints follow the platform naming protocol and carry sidecar files (\texttt{schema.json},
\texttt{ns\_groups.json}, \texttt{train\_config.json}) that inference reads to rebuild the feature schema.

\section{Full upgrade chain}
\label{app:ablations}
Table~\ref{tab:fullchain} reports validation and test AUC for every step of the upgrade chain. Validation AUC
exceeds test AUC by roughly 0.014 at every step, the gap discussed in \S\ref{sec:exp}. The AdagradAR step is
the clearest sign flip: validation AUC falls from 0.84150 to 0.84119 while test AUC rises from 0.826596 to
0.827227.

\begin{table}[htbp]
\centering
\caption{The upgrade chain with validation and test AUC. Steps and folding match Table~\ref{tab:chain}; the
final submission trained without a validation split.}
\label{tab:fullchain}
\small
\begin{tabular}{@{}lrr@{}}
\toprule
Step & Val AUC & Test AUC \\
\midrule
Official baseline (PCVRHyFormer) & 0.82837 & 0.813237 \\
$+$ AMUSE optimizer & 0.82965 & 0.813629 \\
$+$ per-domain time embedding & 0.82960 & 0.814387 \\
$+$ user dense field split & 0.83790 & 0.822375 \\
$+$ item dense split, log1p & 0.83929 & 0.823831 \\
$+$ user raw-stat log1p & 0.84017 & 0.824995 \\
$+$ dense-weight EMA & 0.84074 & 0.825472 \\
$+$ width ($d_\text{model}{=}160$) & 0.84108 & 0.825837 \\
$+$ aligned-pair dense & 0.84139 & 0.826412 \\
$+$ LONGER sequence encoder & 0.84134 & 0.826477 \\
$+$ NS semantic grouping & 0.84150 & 0.826596 \\
$+$ AdagradAR sparse reg. & 0.84119 & 0.827227 \\
$+$ MAM auxiliary head & 0.84136 & 0.827380 \\
$+$ merged single-stream backbone & 0.84175 & 0.827611 \\
$+$ TAPF polarity channel & 0.84197 & 0.827712 \\
$+$ width ($d_\text{model}{=}320$) & 0.84212 & 0.827816 \\
\midrule
Final submission (multi-variable) & n/a & \textbf{0.828535} \\
\bottomrule
\end{tabular}
\end{table}

\end{document}